\documentclass{article}
\usepackage{spconf,amsmath,graphicx}
\usepackage{url}
\usepackage{xcolor}
\usepackage{lineno,multirow}
\usepackage{booktabs}
\usepackage{makecell}
\usepackage[colorlinks,
            linkcolor=blue,       
            anchorcolor=blue,  
            citecolor=blue,        
            ]{hyperref}
\usepackage{enumitem}
\setlist{nosep, leftmargin=14pt}

\usepackage{mwe} 

\title{A SCAN-SPECIFIC UNSUPERVISED METHOD FOR PARALLEL MRI 
RECONSTRUCTION VIA IMPLICIT NEURAL REPRESENTATION}
\name{Ruimin Feng$^{\star}$ \qquad Qing Wu$^{\dagger}$ \qquad Yuyao Zhang$^{\dagger}$ \qquad Hongjiang Wei$^{\star}$}
\address{$^{\star}$School of Biomedical Engineering, Shanghai Jiao Tong University, Shanghai, China; \\
     $^{\dagger}$School of Information and Science and Technology, ShanghaiTech University, Shanghai, China;}
\begin{document}
%
\maketitle
\begin{abstract}
  Parallel imaging is a widely-used technique to accelerate magnetic resonance imaging (MRI). However, current methods still perform poorly in reconstructing artifact-free MRI images from highly undersampled k-space data. Recently, implicit neural representation (INR) has emerged as a new deep learning paradigm for learning the internal continuity of an object. In this study, we adopted INR to parallel MRI reconstruction. The MRI image was modeled as a continuous function of spatial coordinates. This function was parameterized by a neural network and learned directly from the measured k-space itself without additional fully sampled high-quality training data. Benefitting from the powerful continuous representations provided by INR, the proposed method outperforms existing methods by suppressing the aliasing artifacts and noise, especially at higher acceleration rates and smaller sizes of the auto-calibration signals. The high-quality results and scanning specificity make the proposed method hold the potential for further accelerating the data acquisition of parallel MRI.
\end{abstract}
\begin{keywords}
  MRI acceleration, parallel imaging, implicit neural representation, neural network, unsupervised learning
\end{keywords}
\section{Introduction}
\label{sec:intro}
Magnetic Resonance Imaging (MRI) is an important tool in clinical diagnosis. However, the long acquisition time is the main drawback of MRI compared to CT. A variety of methods for reconstructing MRI images from the undersampled k-space data are proposed to reduce the scanning time while maintaining high image quality. Parallel imaging is a widely-used technique that exploits the information redundancy between multiple receiver coils to accelerate the data acquisition of MRI \cite{SENSE,GRAPPA}. Sensitivity encoding (SENSE) \cite{SENSE} and generalized autocalibrating partially parallel acquisition (GRAPPA) \cite{GRAPPA} are the two most popular algorithms of parallel MRI reconstruction. However, these methods suffer from amplified noise at higher acceleration rates due to the ill-posed nature of MRI reconstruction from the undersampled k-space data.

The compressed sensing theory provides an alternative line for MRI acceleration \cite{CS-MRI}, but the requirement of incoherent undersampling patterns hinders its application in routine 2D Cartesian acquisitions. Recently, deep learning has shown great potential in this field. The neural networks are typically trained in a supervised fashion \cite{DL-MRI1,DL-MRI2,DL-MRI3}. Despite impressive results, the large amounts of training data, especially the fully-sampled ones, are unavailable in many practical situations. Therefore, a few scan-specific approaches are further developed, e.g., scan-specific robust artificial-neural-networks for k-space interpolation (RAKI) \cite{RAKI} and a hybrid linear and non-linear approach for scan-specific k-space learning (Residual-RAKI) \cite{Residual-RAKI}. These approaches train nonlinear convolution neural networks on the auto-calibration signal (ACS) region and then use the learned kernel to estimate the missing k-space data. Although these studies outperform the GRAPPA at higher acceleration rates, they still suffer from artifacts when the size of the ACS region is relatively smaller.  

Over the past few years, implicit neural representation (INR) has emerged as a new deep learning paradigm in the field of computer vision \cite{Nerf}. Instead of explicitly performing a specific task, INR models the desired object itself as a continuous function of the spatial coordinates. This representation function is parameterized by the neural network, typically a multilayer perceptron (MLP), and learned under the guidance of the pre-designed loss function. When inferring, the specific task is converted into simply querying the network with the corresponding coordinates. INR has the following advantages: (1) the continuous nature of the representation function enables it to learn the internal redundancies and correlations within an object; (2) it enables easy integration of other explicit prior knowledge and constraints by designing an appropriate architecture.  

Inspired by the new insight of INR, we applied INR to the reconstruction of parallel MRI. Specifically, we modeled the MRI image as a continuous function of the spatial coordinates. This function was parameterized by an MLP. The weights in MLP were optimized by minimizing the loss function containing a data consistency term and a regularization term. Results on different datasets showed that the proposed method outperformed the compared methods, such as GRAPPA, RAKI, and Residual-RAKI, especially at higher acceleration rates and smaller ACS sizes. The main contributions of our study include:
\begin{enumerate}[label={\arabic*)}]
  \item	The proposed method explored the application of INR in parallel MRI reconstruction for the first time.
  \item	The proposed method was scan-specific and unsupervised without involving any external data.
  \item	We proposed an effective framework that could be easily combined with other explicit image regularizers to further improve performance. 
\end{enumerate}

\section{THEORY AND METHODS}
\label{sec:theory&method}

\subsection{Problem Formulation}
\label{ssec:Problem Formulation}
For an MRI imaging system, the undersampled k-space signal of the $j$th channel, $S_j$, is expressed as:
\begin{equation}\label{forward equation}
  S_j = \mathbf{A}_jI+n_j
\end{equation}
where $I$ represents the vectorized image to be reconstructed and $n_j$ is the measurement noise. $\mathbf{A}_j=\mathbf{M}\mathbf{F}\mathbf{C}_j$ is the forward physical model, where $\mathbf{C}_j$ denotes the diagonalized sensitivity map matrix of the $j$th coil, $\mathbf{F}$ denotes the Fourier transform and $\mathbf{M}$ is the diagonalized sampling mask. In traditional reconstruction algorithms, the image $I$ is restored by minimizing the following objective function:   
\begin{equation}\label{opt function1}
  \mathop{\arg\min}\limits_{I}{\frac{1}{2} \sum_{j = 1}^{c}  \Vert S_j-\mathbf{A}_jI \Vert_2^2 + \lambda \mathcal{R} (I)}
\end{equation}
where $c$ is the number of channels. The first term represents data consistency with the measured signals. $\mathcal{R} (I)$ is the regularization term that imposes prior information on the reconstructed image $I$ and $\lambda$ is a tunable parameter that balances the contributions of these two terms.   

From another perspective, the image intensities can be regarded as a continuous function of the spatial coordinates, i.e., $f_\theta (x,y)$, where the subscript $\theta$ indicates the parameters of this function and $(x,y)$ represents the coordinates in two spatial dimensions. Let $I_\theta$ be the discretized image matrix after uniformly sampling $f_\theta$ at the pixel locations, then the MRI reconstruction problem is rewritten as:
\begin{equation}\label{opt function2}
  \mathop{\arg\min}\limits_{\theta }{\frac{1}{2} \sum_{j = 1}^{c}  \Vert S_j-\mathbf{A}_jI_\theta  \Vert_2^2 + \lambda \mathcal{R} (I_\theta )}
\end{equation}
Thus, this problem has been transformed into optimizing the parameters of the continuous function, instead of directly operating on the desired maps.

In the field of INR, the fully-connected neural network, i.e., MLP, is used to parameterize $f_\theta$. However, previous studies have shown that a pure MLP is biased towards lower-frequency functions and poorly represents high-frequency information \cite{spectral-bias}. To overcome this problem, we adopted the sinusoidal representation network (SIREN) \cite{SIREN} to better represent the details and fine structures of an image. Specifically, SIREN leverages the periodic sine function to replace the ReLU activation function in MLP. Moreover, SIREN requires a specific initialization scheme for effective training, where a hyperparameter $w_0$ is involved. $w_0$ controls the spatial frequency of the first layer, which determines the network's ability to represent high-frequency information. In this study, $w_0$ is scan-dependent that needs to be fine-tuned for different datasets.    
\subsection{Overall framework of the proposed method}
\label{ssec:Overall framework of the proposed method}
\begin{figure*}[htb]
  \centering
  \includegraphics[width=150mm]{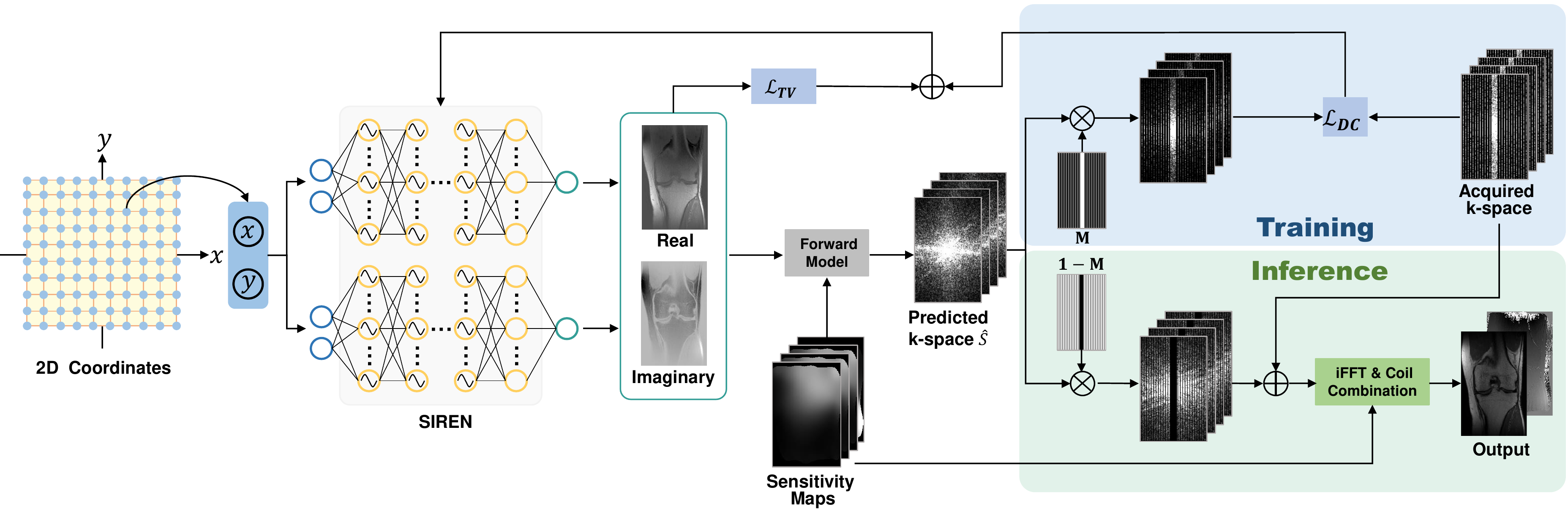}
  \caption{
    Overview of the proposed method. The spatial coordinates of the image were fed into two SIREN networks to output the corresponding real and imaginary intensities, respectively. Then, the output intensities were used to predict the k-space signal $\widehat{S}$. The network weights were optimized by minimizing the data consistency loss $\mathcal{L}_{DC}$ and the total variation loss $\mathcal{L}_{TV}$. When inferring, the acquired k-space data replaced the predicted k-space to impose data consistency, and these composite data were used to reconstruct the final image. 
    }
  \label{fig1}
\end{figure*}

Fig. \ref{fig1} illustrates the overview of the proposed method. First, spatial coordinates of the image were fed into two SIREN networks to output the corresponding real and imaginary intensities, respectively. Then, the output intensities were used to predict the k-space signal $\widehat{S}$ via the forward physical model, where the coil sensitivity maps were pre-calculated using the ESPIRiT algorithm \cite{ESPIRiT}. During training, the network weights were optimized by minimizing the following loss function through the Adam optimizer \cite{Adam}:
\begin{equation}\label{loss function tot}
   \mathcal{L}_{tot} =\mathcal{L}_{DC}+\lambda \mathcal{L}_{TV} 
\end{equation}  
where $\mathcal{L}_{DC}$ imposes data consistency with the acquired k-space data, $\mathcal{L}_{TV}$ represents the total variation regularization term:
\begin{equation}\label{loss function DC}
  \mathcal{L}_{DC}=\sum_{j = 1}^{c} \Vert S_j-\mathbf{A}_jI_\theta  \Vert_1 
\end{equation}      
\begin{equation}\label{loss function TV}
  \mathcal{L}_{TV}=\Vert G(I_\theta)  \Vert_1 
\end{equation} 
where $G$ is the gradient operator.

When inferring, the acquired k-space data replaced the predicted k-space to enforce data consistency. Finally, the individual coil images were obtained by applying the 2D inverse Fourier transform to these composite k-space data and was combined using the SENSE-1 method \cite{SENSE}. 

\subsection{Implementation details}
\label{ssec:Implementation details}
The proposed method was implemented using PyTorch 1.10.2 in Python 3.9 on a workstation with two Intel Xeon Platinum 8249C CPUs @ 2.10GHz with 256 GB RAM and an NVIDIA GeForce RTX 3090 GPU with 24 GB memory.
\section{EXPERIMENTS}
\label{sec:EXPERIMENTS}

\subsection{Datasets}
\label{ssec:Datasets}
Two public datasets were used in this study. One is a single-slice fully-sampled brain k-space from the study \cite{SPARK}, which was acquired using the MPRAGE sequence with 32 receiver channels. Another dataset is a fully-sampled 15-channel knee k-space, obtained from the NYU fastMRI Initiative database \cite{fastMRI1}. This dataset was acquired using the 2D turbo spin-echo sequence without fat suppression. The hyperparameter settings when reconstructing these datasets are listed in Table \ref{table1}.
\begin{figure}[h]
  \centering
  \includegraphics[width=80mm]{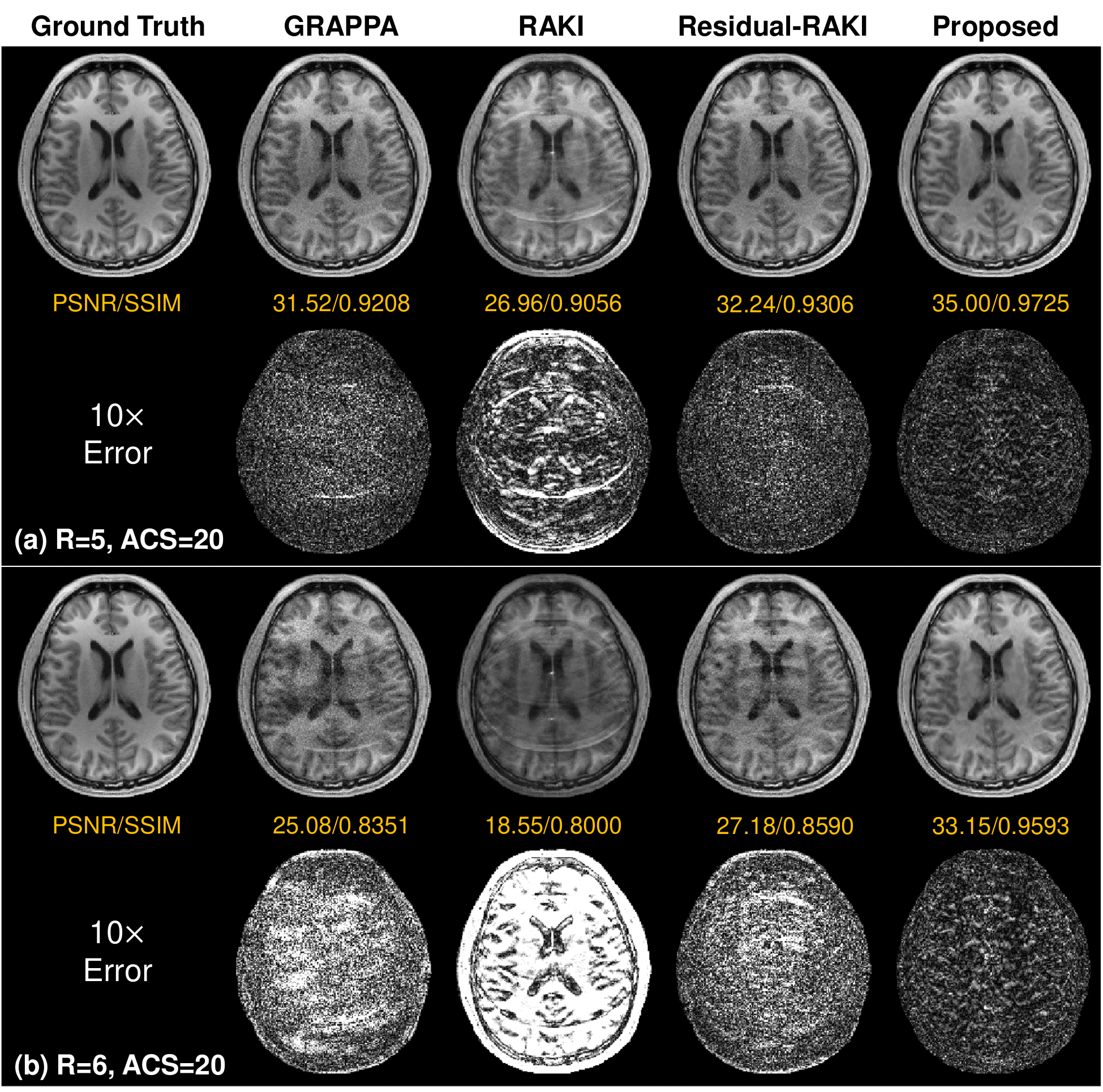}
  \caption{
    Comparisons between GRAPPA \cite{GRAPPA}, RAKI \cite{RAKI}, Residual-RAKI \cite{Residual-RAKI}, and the proposed method on the 32-channel brain dataset. 
    }
  \label{fig2}
\end{figure}
\begin{table}[htbp]
  \centering
  \caption{List of hyperparameter settings.}
    \begin{tabular}{ccc}
    \hline
    \textbf{Hyperparameters} & \makecell[c]{\textbf{Brain} \\ \textbf{Dataset}} & \makecell[c]{\textbf{Knee} \\ \textbf{Dataset}} \\
    \hline
    $w_0$ & 10    & 25 \\
    TV loss weight  & 0.05  & 3 \\
    Number of layers in SIREN & 10    & 8 \\
    Number of hidden neurons & 256   & 256 \\
    Number of iterations & 4000  & 4000 \\
    Learning rate & 1e-4 & 1e-4 \\
    \hline
    \end{tabular}%
  \label{table1}%
\end{table}%
\subsection{Performance evaluation}
\label{ssec:Performance evaluation}
The proposed method was compared with the GRAPPA algorithm \cite{GRAPPA}, and two scan-specific methods, RAKI \cite{RAKI} and Residual-RAKI \cite{Residual-RAKI}. To evaluate the performance of the proposed method, the datasets were retrospectively uniformly undersampled with different acceleration rates and ACS sizes (R=\{5, 6\} and ACS=\{20, 24, 30, 36, 40\} for the brain dataset,  R=\{4, 5\} and ACS=\{24, 32\} for the knee dataset). The peak signal-to-noise ratio (PSNR) and structural similarity index (SSIM) were calculated for quantitative evaluation.
\subsection{Ablation study}
\label{ssec:Ablation study}
Ablation experiments were conducted to illustrate the effectiveness of each configuration of the proposed method. Different variants were designed by using the traditional MLP (i.e., with ReLU activation function) to replace the SIREN network (termed Proposed-Sine), ablating the TV loss function (termed Proposed-TV), and ablating the k-space consistency step in inference (termed Proposed-KC). These models were compared on the knee dataset with different acceleration rates and ACS sizes.
\begin{table*}[htbp]
  \begin{center}
    \caption{Quantitative evaluation metrics of different variants in the ablation experiments on the 15-channel knee dataset.}
    \label{table2}
    \begin{tabular}{cccccc}
    \hline
    \textbf{ACS size} & \textbf{Acceleration rate} & \textbf{Proposed-Sine} & \textbf{Proposed-TV} & \textbf{Proposed-KC} & \textbf{Proposed} \\
    \hline
    \multirow{2}{*}{ACS=24} & R=4   & 36.22/0.9001 & 37.64/0.9171 & 39.09/0.9201 & \textbf{39.48/0.9289} \\
         & R=5   & 36.02/0.8992 & 36.25/0.9033 & 38.37/0.9194 & \textbf{38.76/0.9238} \\
    \hline
    \multirow{2}{*}{ACS=32} & R=4   & 36.47/0.9038 & 37.98/0.9212 & 38.97/0.9219 & \textbf{39.49/0.9321} \\
        & R=5   & 35.82/0.9011 & 36.89/0.9132 & 38.49/0.9202 & \textbf{39.04/0.9296} \\
    \hline
    \end{tabular}%
  \end{center}
\end{table*}%
\begin{figure}[h]
  \centering
  \includegraphics[width=80mm]{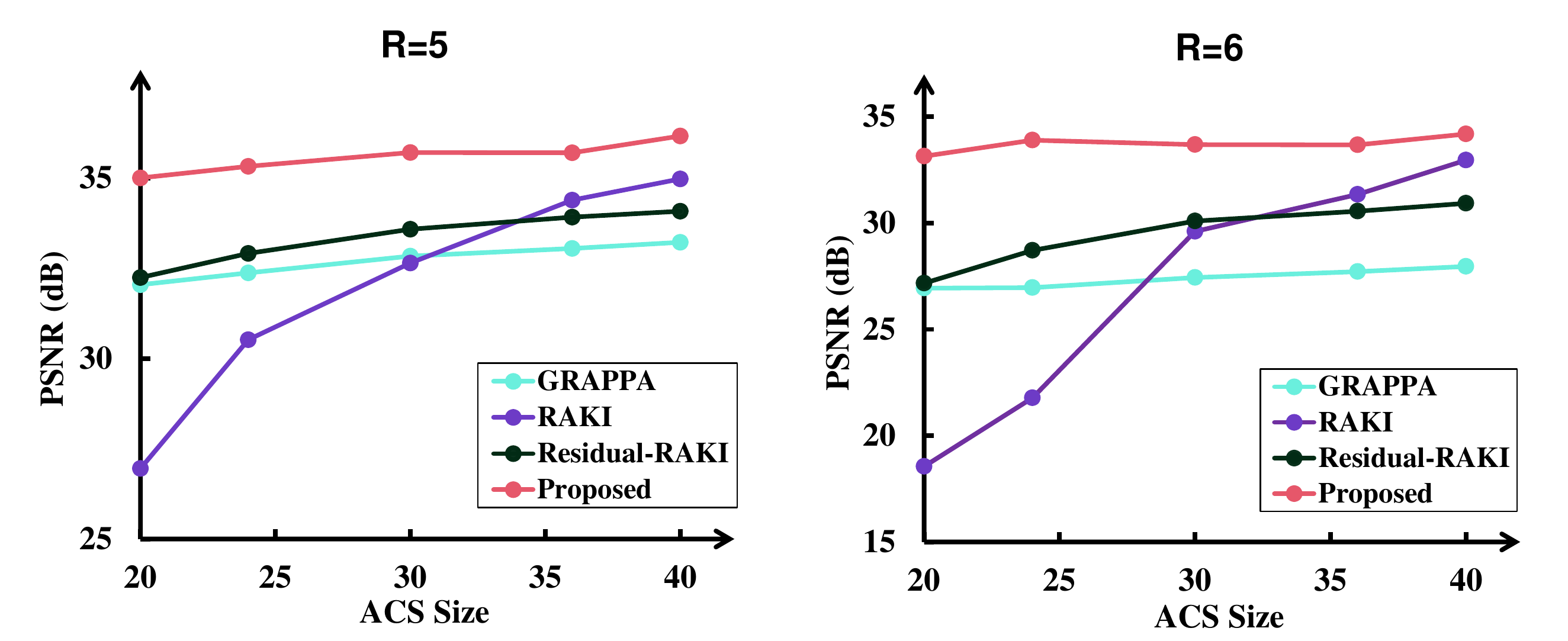}
  \caption{
    Performance variations of GRAPPA \cite{GRAPPA}, RAKI \cite{RAKI}, Residual-RAKI \cite{Residual-RAKI}, and the proposed method on the brain dataset as a function of ACS sizes at R=5 and R=6, respectively.
    }
  \label{fig3}
\end{figure}
\begin{figure}[h]
  \centering
  \includegraphics[width=80mm]{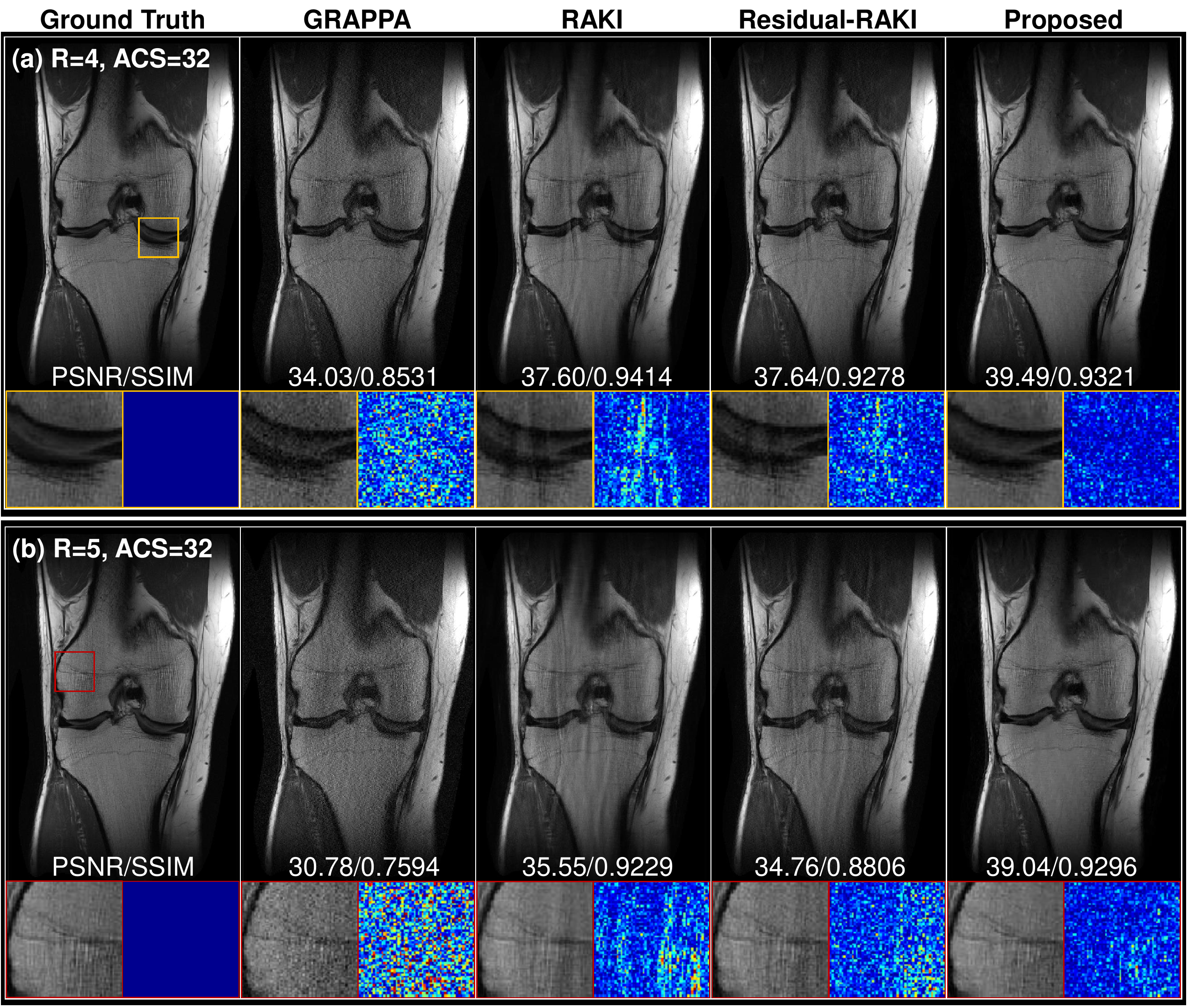}
  \caption{
    Comparisons between GRAPPA \cite{GRAPPA}, RAKI \cite{RAKI}, Residual-RAKI \cite{Residual-RAKI}, and the proposed method on the 15-channel knee dataset. 
    }
  \label{fig4}
\end{figure}
\section{RESULTS}
\label{sec:RESULTS}
\subsection{Comparison with baseline methods}
\label{ssec:Comparison with baseline methods}
Fig. \ref{fig2} shows the reconstruction results on the 32-channel brain data with only 20 ACS lines. The proposed method produces improved results visually. In contrast, GRAPPA, RAKI, and Residual-RAKI suffer from artifacts and noise. Quantitatively, the proposed method achieves the highest PSNR of 35.00 dB at R=5 and 33.15 dB at R=6 and the highest SSIM of 0.9725 at R=5 and 0.9593 at R=6. Fig. \ref{fig3} plots the performance variation of these methods as a function of the number of ACS lines. The proposed method is more robust to the ACS size and outperforms the compared methods, particularly at smaller ACS sizes.

Fig. \ref{fig4} presents the results of the 15-channel knee data. At R=4, the proposed method successfully removes artifacts and noise that are apparent in the results of the compared methods, as illustrated by the zoomed-in images in Fig. \ref{fig4}(a). Similar results are observed when the acceleration rate reaches 5 (Fig. \ref{fig4}(b)). The proposed method exhibits comparable performances when applied to the undersampled data at R=4 and R=5 (39.49 dB/0.9321 \textit{vs} 39.04 dB/0.9296). In contrast, the performance of the compared methods degrade at R=5. Notably, the proposed method achieves a PSNR of 39.49 dB, nearly 2 dB higher than Residual-RAKI, and an SSIM of 0.9321 at R=4. For R=5, the proposed method obtains a PSNR of 39.04, more than 4 dB higher than Residual-RAKI, and an SSIM of 0.9296. 
\subsection{Time consumption}
\label{ssec:Time consumption}
The consumption times for RAKI, Residual-RAKI, and the proposed method are 2.2 min, 8.5 min, and 2.5 min on the brain dataset and 1.5 min, 5.1 min, and 7.8 min on the knee dataset. Therefore, the proposed method takes comparable time with RAKI and Residual-RAKI while achieving superior reconstruction results.
\subsection{Results of the ablation experiments}
\label{ssec:Results of the ablation experiments}
Table \ref{table2} lists the quantitative evaluations obtained by the variants described in Section \ref{ssec:Ablation study}. It can be seen that the proposed full model achieves the best performance at different acceleration rates and ACS sizes, demonstrating the effectiveness of the current configurations.
\section{CONCLUSION}
\label{sec:CONCLUSION}
In this study, we proposed a novel scan-specific framework for parallel MRI reconstruction based on INR. Instead of directly performing the reconstruction task, the proposed method alternatively models the MRI image as a continuous function of spatial coordinates. Benefitting from the synergy of the implicit continuous property provided by INR and the explicit regularizer, the proposed method shows superior performance on different datasets with various acceleration rates and ACS sizes. The high-quality results yielded by the proposed method and the scan-specific characteristics make it potential for further accelerating MRI data acquisition.

\section{Compliance with ethical standards}
\label{sec:ethics}
This research study was conducted retrospectively using human subject data made available in open access by \url{https://github.com/YaminArefeen/spark\_mrm\_2021} for the brain dataset and \url{https://fastmri.org/dataset/} for the knee dataset. Ethical approval was not required as confirmed by the license attached with the open access data.

\section{Acknowledgments}
\label{sec:acknowledgments}
This work is supported by the National Natural Science Foundation of China (91949120, 61901256). The authors have no relevant financial or non-financial interests to disclose. 


\bibliographystyle{IEEEbib}
\bibliography{MRI_recon}
\end{document}